\documentclass{emulateapj} %{aastex} 

\usepackage{epstopdf} 
\usepackage[usenames]{color} 
 
\newcommand{\Teff}{T_{\rm eff}} 

\newcommand{\kepler}{\emph{Kepler}}
\newcommand{\corot}{\emph{CoRoT}}

\shorttitle{Hybrid $\gamma$\,Dor -- $\delta$\,Sct pulsators from {\kepler}} 
\shortauthors{Grigahc\`ene et al.} 
 
\begin{document} 
 
\title{Hybrid $\gamma$\,Doradus -- $\delta$\,Scuti pulsators: \\[5pt] 
 New insights into the physics of the oscillations from {\kepler} 
observations\\[5pt]} 

\author{A.~Grigahc\`ene\altaffilmark{1}, 
   V.~Antoci\altaffilmark{2},
   L.~Balona\altaffilmark{3},
   G.~Catanzaro\altaffilmark{4},
   J.~Daszy{\'n}ska-Daszkiewicz\altaffilmark{5},  
   J.~A.~Guzik\altaffilmark{6}, 
   G.~Handler\altaffilmark{2}, 
   G.~Houdek\altaffilmark{2},
   D.~W.~Kurtz\altaffilmark{7},
   M.~Marconi\altaffilmark{8},
   M.~J.~P.~F.~G.~Monteiro\altaffilmark{1},
   A.~Moya\altaffilmark{9},
   V.~Ripepi\altaffilmark{8},
   J.-C.~Su\'arez\altaffilmark{10},
   K.~Uytterhoeven\altaffilmark{11},
   W.~J.~Borucki\altaffilmark{12},
   T.~M.~Brown\altaffilmark{13},
   J.~Christensen-Dalsgaard\altaffilmark{14},
   R.~L.~Gilliland\altaffilmark{15},
   J.~M.~Jenkins\altaffilmark{16},
   H.~Kjeldsen\altaffilmark{14},
   D.~Koch\altaffilmark{12},
   S.~Bernabei\altaffilmark{17},
   P.~Bradley\altaffilmark{18},
   M.~Breger\altaffilmark{2},
   M.~Di~Criscienzo\altaffilmark{19},
   M.-A.~Dupret\altaffilmark{20},
   R.~A.~Garc{\'i}a\altaffilmark{11},
   A.~Garc\'{i}a~Hern\'{a}ndez\altaffilmark{10},
   J.~Jackiewicz\altaffilmark{21},
   A.~Kaiser\altaffilmark{2},
   H.~Lehmann\altaffilmark{22},
   S.~Mart\'{\i}n-Ruiz\altaffilmark{10},
   P.~Mathias\altaffilmark{23},
   J.~Molenda-\.Zakowicz\altaffilmark{5},
   J.~M.~Nemec\altaffilmark{24},
   J.~Nuspl\altaffilmark{25},
   M.~Papar{\'o}\altaffilmark{25},
   M.~Roth\altaffilmark{25}, 
   R.~Szab\'o\altaffilmark{25},
   M.D.~Suran\altaffilmark{27},
   R.~Ventura\altaffilmark{4}
} 
 
\altaffiltext{1}{Centro de Astrof\'{\i}sica, Faculdade de Ci\^encias, Universidade do Porto, Rua das Estrelas, 4150-762 Porto, Portugal} 
%\email{ahmed.grighacene@astro.up.pt} 
\altaffiltext{2}{Institut fuer Astronomie, Tuerkenschanzstr. 17, A-1180 Wien, Austria}
\altaffiltext{3}{South African Astronomical Observatory, P.O. Box 9, Observatory 7935, South Africa}
\altaffiltext{4}{INAF - Osservatorio Astrofisico di Catania, Via S. Sofia 78, 95123 Catania, Italy}
\altaffiltext{5}{Instytut Astronomiczny, Uniwersytet Wroc{\l}awski, Kopernika 11, 51-622 Wroc{\l}aw, Poland}
\altaffiltext{6}{Los Alamos National Laboratory, X-2 MS T-086, Los Alamos, NM 87545-2345, USA}
\altaffiltext{7}{Jeremiah Horrocks Institute of Astrophysics, University of Central Lancashire, Preston PR1\,2HE, UK}
\altaffiltext{8}{INAF - Osservatorio Astronomico di Capodimonte, Via Moiariello 16, 80131 Naples, Italy}
\altaffiltext{9}{Laboratorio de Astrof\'isica Estelar y Exoplanetas, LAEX-CAB (INTA-CSIC), PO BOX 78, 28691 Villanueva de la Ca\~nada, Madrid, Spain}
\altaffiltext{10}{Instituto de Astrof\'{i}sica de Andaluc\'{i}a (CSIC), CP3004, Granada, Spain}
\altaffiltext{11}{Laboratoire AIM, CEA/DSM CNRS - U. Paris Diderot, IRFU/SAp, Centre de Saclay, 91191 Gif-sur-Yvette Cedex, France}
\altaffiltext{12}{NASA Ames Research Center, MS 244-30, Moffett Field, CA 94035, USA}
\altaffiltext{13}{Las Cumbres Observatory Global Telescope, Goleta, CA 93117, USA}
\altaffiltext{14}{Department of Physics and Astronomy, Aarhus University, DK-8000 Aarhus C, Denmark}
\altaffiltext{15}{Space Telescope Science Institute, Baltimore, MD 21218, USA}
\altaffiltext{16}{SETI Institute/NASA Ames Research Center, MS 244-30, Moffett Field, CA 94035, USA}
\altaffiltext{17}{INAF - Osservatorio Astronomico di Bologna, Via Ranzani 1, 40127 Bologna, Italy}
\altaffiltext{18}{Los Alamos National Laboratory, X-4 MS T-087, Los Alamos, NM 87545-2345, USA}
\altaffiltext{19}{INAF - Osservatorio Astronomico di Roma, via Frascati 33, 00040 Monte Porzio Catone, Rome, Italy}
\altaffiltext{20}{Institut d'Astrophysique et de G\'eophysique, Universit\'e de Li\`ege, All\'ee du 6 Ao{\^u}t 17-B 4000 Li\`ege, Belgium}
\altaffiltext{21}{Department of Astronomy, New Mexico State University, Las Cruces, NM 88001, USA}
\altaffiltext{22}{Thueringer Landessternwarte, 07778 Tautenburg, Germany}
\altaffiltext{23}{Dpt Fizeau, UMR 6525 Observatoire de la C{\"o}te d'Azur/CNRS, BP 4229, F06304 Nice Cedex 4, France}
\altaffiltext{24}{Deptartment of Physics \& Astronomy, Camosun College, Victoria, British Columbia, Canada}
\altaffiltext{25}{Konkoly Observatory of the Hungarian Academy of Sciences, P.O. Box 67, H-1525 Budapest, Hungary}
\altaffiltext{26}{Kiepenheuer Institut f\"ur Sonnenphysik, Sch\"oneckstr. 6, 79104, Freiburg, Germany}
\altaffiltext{27}{Astronomical Institute of the Romanian Academy, Str. Cutitul de Argint 5, RO 40557, Bucharest, Romenia}

\begin{abstract}
Observations of the pulsations of stars can be used to infer their interior structure and test theoretical models.
The main sequence $\gamma$\,Doradus (Dor) and $\delta$\,Scuti (Sct) stars with masses 1.2-2.5~$M_{\sun}$ are particularly useful for these studies.
The $\gamma$\,Dor stars pulsate in high-order $g$~modes with periods of order 1 day, driven by convective blocking at the base of their envelope convection zone.
The $\delta$\,Sct stars pulsate in low-order $g$ and $p$~modes with periods of order 2 hours, driven by the $\kappa$ mechanism operating in the He\,\textsc{ii} ionization zone.
Theory predicts an overlap region in the Hertzsprung-Russell diagram between instability regions, where 'hybrid' stars pulsating 
in both types of modes should exist. 
The two types of modes with properties governed by different portions of the stellar interior provide complementary model constraints.
Among the known $\gamma$\,Dor and $\delta$\,Sct stars, only four have been confirmed as hybrids.
Now, analysis of combined Quarter 0 and Quarter 1 {\kepler} data for hundreds of variable stars shows that the frequency spectra are so rich that there are practically no pure $\delta$\,Sct or $\gamma$\,Dor pulsators, i.e. essentially all of the stars show frequencies in both the
$\delta$\,Sct and $\gamma$\,Dor frequency range.  A new observational classification scheme is proposed that takes into account the amplitude as well as the frequency, and is applied to categorize 234 stars as $\delta$\,Sct, $\gamma$\,Dor, $\delta$\,Sct/$\gamma$\,Dor or $\gamma$\,Dor/$\delta$\,Sct hybrids.
\end{abstract} 


%-------------------------------------------------------------------------
\section{Introduction}

Many stars oscillate in multiple simultaneous modes during some portion of their evolution.
The frequency range, spacing, amplitude, or other properties can be used to infer the interior structure of these stars and test the input physics and predictions of theoretical models, a process called asteroseismology. 
The $\gamma$\,Doradus (Dor) and $\delta$\,Scuti (Sct) stars that pulsate in many simultaneous frequencies are particularly useful for asteroseismic studies.
These stars are core hydrogen-burning and somewhat more massive than the Sun, at 1.2-2.5~$M_{\sun}$. However, unlike the Sun, they have convective cores, shallower convective envelopes, and often rapid rotation.

While helioseismologists have been searching for solar $g$~modes (e.g., \citealt{Garc07}), stellar astronomers have found them in several groups of pulsating stars including the $\gamma$\,Dor variables. The $\gamma$\,Dor $g$~modes have periods of the order of one day, making them difficult to observe continuously from the ground. In addition, the $g$~modes reach the stellar surface with small amplitudes which make their detection a challenge.

From a pulsational point of view, there is a clear distinction between $\delta$\,Sct and $\gamma$\,Dor stars.
The former are short-period pulsating stars, with periods between 0.014 and 0.333 \,day (d), while the latter are long-period pulsators, with periods between about 0.3 and 3\,d.
The distinction is even clearer if one considers the pulsation constant $Q$ \citep{Hand02}; the $\delta$\,Sct stars have $Q < 0.055$\,d whereas for $\gamma$\,Dor stars $Q > 0.24$\,d.
When the $\gamma$\,Dor variables were first recognized as a new group \citep{Balona94, Kaye99}, their position on the Hertzsprung-Russell diagram (HRD) already suggested some relationship with the $\delta$\,Sct group.
The $\gamma$\,Dor stars lie in a zone close to the cool border of the classical instability strip, partially overlapping with the $\delta$\,Sct pulsators.
This latter property led astronomers to investigate the possible existence of hybrid stars exhibiting both types of pulsations \citep{Breg96}.

These hybrid objects are of great interest because they offer additional constraints on stellar structure.
The $\gamma$\,Dor stars pulsate in $g$~modes, giving us the opportunity to probe the stellar core, while the $\delta$\,Sct stars pulsate in $p$~modes, that help to probe the stellar envelope.
Moreover, since $\gamma$\,Dor stars pulsate in high-order $g$~modes, the asymptotic approximation can be used, where we can benefit from an analytical solution of the equations \citep{Smey07}.
Hence, the existence of hybrid stars provides a unique opportunity to 
extend this advantage to $\delta$\,Sct stars \citep{Hand02}.

Furthermore, in the same region of the HRD where the $\delta$\,Sct and $\gamma$\,Dor instability strips overlap, solar-like oscillations are also predicted to occur for $\delta$\,Sct stars \citep{Houd99, Samadi02}. So far, no detections of solar-like oscillations in such hot stars have been reported \citep{Antoci09}, which is probably due to the detection limits imposed by earthbound measurements. Stochastic excitation as an excitation mechanism in $\gamma$\,Dor stars has also been searched for (e.g. see \citealt{Pereira07}), but so far not confirmed.
{\kepler} data are expected to reach the precision needed to observe stochastically excited pulsation.
A positive or a significant null detection would set strong constraints on the parametrization of the convective envelopes.

\citet{Poretti09}, using data from the French-led {\corot} mission, show that over most of the periodogram of HD~292790 (V = 9.5), the amplitude of the noise level is 0.01 mmag for an observational time span of 57 d. In {\kepler} observations of the combined Quarter 0 and Quarter 1 data of a 9.5 mag star (time span of 44 d) the amplitude of the noise level is 0.001 mmag.  The noise level in {\kepler} observations is therefore approximately 10 times smaller than in {\corot} observations.

The large number of stars to be studied by {\kepler} will also allow for control of long-term (low-frequency) instrumental changes; understanding these is important for confident identification of the low-frequency $g$~modes.
Most importantly, the 3.5-y time span of the mission will give data with unprecedented frequency resolution.
This is known to be necessary from ground-based studies that show closely spaced frequencies in data sets spanning years. 

%-------------------------------------------------------------------------
\section{Current observational status}
%*************
\begin{figure}
\epsscale{1.1}
\plotone{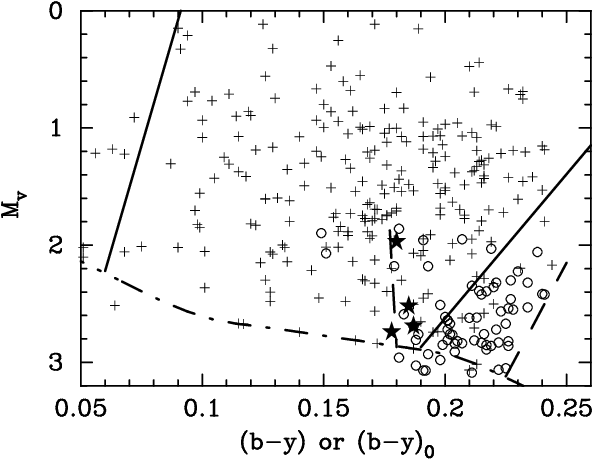}
\caption{Color-Magnitude diagram of $\delta$ Sct (plus signs, \citealt{Rod00}), $\gamma$ Dor (open circles, \citealt{Hen07}) and stars that
show both types of pulsation (star symbols, \citealt{Hen05, Rowe06, Uytt08, Hand09}, Rowe, priv. comm.).
The Zero-Age Main Sequence (dashed-dotted line, \citealt{Crawford79}) as well as
the observed borders of the $\delta$ Sct (solid lines, \citealt{Rod01}) and $\gamma$ Dor (dashed lines, \citealt{Hand02}) instability strips, are indicated.
Lines on the left indicate blue edges while those on the right give the red edges of the instability strips.
}
 \label{fig:hrdiag}
\end{figure}

Among the hundreds of known $\delta$\,Sct stars and several dozen confirmed
$\gamma$\,Dor variables, only 4 have been previously observationally identified as hybrids 
 and interpreted afterward using theoretical models \citep{Boua09}.

One of the major contributions came from the Canadian Space Mission \emph{MOST} \citep{Matt07}, while {\corot} \citep{Auve09} is expected to add new confirmed cases in the near future, especially since the preliminary results indicate the possible detection of hybrid $\gamma$\,Dor and $\delta$\,Sct pulsators in the exofield \citep{Math09}.

The known hybrid pulsators are depicted, together with single-type pulsators, in Fig.~\ref{fig:hrdiag}.
The first case recorded was \objectname{HD~209295} \citep{Hand02}, but it turned out afterward that this star is a close binary and the long period modes could be tidally excited \citep{Hand02b}.
Following this work, \citet{Hen05} discovered both $\gamma$\,Dor and $\delta$\,Sct-type pulsations in the single Am star \objectname{HD~8801} \citep{Hand09}.

\emph{MOST} found two additional hybrid pulsators: \objectname{HD~114839} \citep{King06} and \objectname{BD+18~4914} \citep{Rowe06}.
Both of these are also Am stars.
In preparation for the {\corot} mission, \citet{Uytt08} identified \objectname{HD~49434} as a candidate hybrid pulsator.

It is very interesting to note that three of the four known hybrids show Am type 
peculiarities, even though Am stars are less likely to show pulsational 
variability than A/F-type main-sequence stars \citep{Kurtz89}.
Increasing the number of hybrids is expected to shed light on the relationship between hybrid pulsation and Am chemical peculiarity.

%-------------------------------------------------------------------------
\section{Theoretical expectations and modeling}

%*************
\begin{figure}
\epsscale{1.1}
\plotone{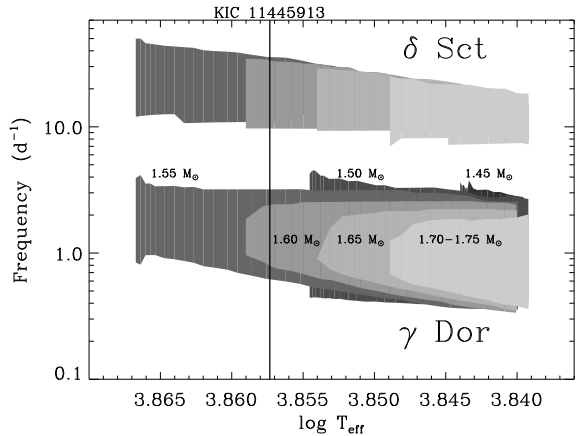}
\caption{Hybrid theoretical instability domain for $\ell=0-3$ $g$ and $p$~modes.
Models have: $Z=0.02$, $\alpha_{\rm MLT}=2.0$, $\alpha_{ov}=0.2$. The vertical line corresponds to the effective temperature of the \,{\kepler} target: KIC 11445913 ($\Teff=7200 \pm 200$ K).
\label{fig:instreg}}
\end{figure}

It is well known that the $\kappa$-mechanism operating in the He\,\textsc{ii} ionization zone is responsible for the pulsational driving of 
$\delta$\,Sct stars \citep{Chev71}.
With this mechanism, the theoretical blue edge of the $\delta$\,Sct instability strip is predicted correctly with linear non-adiabatic models for radial as well as for non-radial modes.
But determining the theoretical red edge is more difficult, because it requires a time-dependent treatment of the 
interaction between convection and pulsation.
In this stage of their evolution, cool $\delta$\,Sct stars have 
relatively deep surface convection zones which typically extend beyond the 
location of the second stage of helium ionization, the principal driving region of $\delta$\,Sct pulsations.
Consequently, mode stability is crucially affected by 
convection dynamics for stars located near the red edge of the classical 
instability strip in the HRD.
The location of the red edge can be described theoretically only by means of 
a time-dependent treatment of the turbulent fluxes (e.g., 
\citealt{Deupr77,Bak79,Gnz82,Stel84}). Moreover, the observed location of the instability strip provides independent information for calibrating convection models.

Several authors (e.g., \citealt{Bon99, Houd00, Dupt05, Xng07}) have 
successfully modeled the red edge of the classical instability strip but reported that different physical mechanisms are responsible for stabilizing the pulsations. 
It is therefore of great importance that our calculations for mode stability be reassessed and {\kepler} is in the position to provide the necessary high-quality data for calibrating and then testing the models for convection in classical pulsators. 
Moreover, analysis of the follow-up ground-based
multi-color photometry can yield valuable constraints
on efficiency of the convective transport in these pulsators \citep{Dasz03}.

\citet{Guzik00}, using frozen-convection models, and \citet{Dupt05,Dupt06}, using time-dependent convection \citep{Grig05}, have shown that the position of the convective envelope base is the key to driving $\gamma$\,Dor $g$~modes.
The convective blocking mechanism, operating when the convective timescale becomes comparable to the $g$~mode period, can explain the location of the blue edge of the $\gamma$\,Dor instability strip and its sensitivity to the mixing length parameter $\alpha_{MLT}$.
On the other hand, the red edge of the $\gamma$\,Dor instability strip is caused by
radiative damping in the $g$~mode cavity, which dominates over the excitation near the 
convective envelope base.
As an example, we show in Fig.~\ref{fig:instreg} the hybrid theoretical instability domain for $\ell=0-3$ $g$ and $p$~modes. For the calculated grid ($Z{=}0.02$, $\alpha_{\rm MLT}{=}2.0$), the effective temperature range of hybrids is 6900~K $\lesssim$ $\Teff$ $\lesssim$ 7500~K.
The vertical line in Fig.~\ref{fig:instreg} represents a model of a hybrid pulsator for which we have unstable high order $g$~modes, corresponding to $\gamma$\,Dor range as well as unstable low order $p$ and $g$~modes, corresponding to typical $\delta$\,Sct pulsations separated by a gap of stable modes.

We also note that the stability of modes in the gap decreases as degree $\ell$ 
increases.
This is due to the fact that the Lamb frequency $S^{2}_{\ell} \propto \ell (\ell+1)$, so the size of the propagation cavity, of $g$~modes increases with $\ell$ for a given frequency.

Therefore, the direct comparison of the observed frequency spectrum with models is a test of the excitation models and can improve our knowledge about important physical characteristics of the stars, including the age, chemical composition, etc.

Another more general problem in stellar pulsation modeling is connected with the effects of rotation on eigenfrequencies and eigenfunctions.
The commonly adopted approach is perturbation theory which assumes that oscillation frequencies are much larger than the angular rotation rate $\Omega$ and that the star is not significantly deformed by the centrifugal force ($\Omega \ll \Omega_{crit}\equiv\sqrt{G M/R^{3}}$) \citep{Reese08}. This assumption is violated for many $\delta$\,Sct and $\gamma$\,Dor stars, due to their relatively rapid rotation. This implies that rotation will significantly affect the pulsation frequency spectrum and surface amplitude distribution. This in turn will affect how limb darkening and light cancellation affects the pulsation mode visibility and radial velocity variations.

Hybrids can be extremely useful when investigating the internal rotation profile of stars.
In \citet{Suarez06} the authors examine the effect of rotation on the oscillation spectrum for intermediate-mass rotating stars. Significant effects for both $p$ and mixed modes (very low radial order $g$ and $p$~modes) are predicted.
Analysis of hybrid stars will allow us to constrain the  modeling of their rotation profiles. Better constraints on the internal rotation profile will help us better understand angular momentum transport processes in the stellar interior.

Neither $\delta$\,Sct stars nor hybrid stars are expected to pulsate in asymptotic $p$~mode oscillations.
However, recently \citet{Garch09} found regular spacings in the oscillation spectrum of the {\corot} $\delta$\,Sct star HD174936. Analysis of the 422 detected oscillation frequencies revealed a spacing periodicity of around 52~$\mu{\rm Hz}$.
Although the modes considered were not in the asymptotic regime, a comparison with stellar models confirmed that this signature may stem from a quasi-periodic pattern similar to the so-called large separation in solar-like stars.
Other frequency spacings have also been observed in other stars using ground based observations (e.g. \citealt{Hand00, Breger09}).
From numerical results, this approximate asymptotic behaviour has been found to hold for p modes of low order  (e.g. \citealt{Jcd00}). 
The presence of such spacings in hybrid stars may allow us to analyze simultaneously $p$ and $g$~modes with some characteristics of the asymptotic regime and even solar-like oscillations, if detected. 

Interestingly, asteroseismology of hybrid stars can also benefit  from techniques used for analyzing and interpreting the oscillation spectra of both type of pulsations.
For instance, if $g$~modes in the asymptotic regime are detected, techniques based on the asymptotic properties of $g$~modes can be applied, namely the Frequency Ratio Method (FRM -- \citealt{Moya05, Suarez05}).
The FRM is particularly adapted for obtaining asteroseismic information on $\gamma$\,Dor pulsating stars showing at least three pulsation frequencies. 
The method provides an identification of the radial order $n$ and degree $\ell$ of observed frequencies and an estimate of the integral of the buoyancy frequency (Brunt-V\"ais\"al\"a) weighted over the stellar radius along the radiative zone.
\citet{Migl08} also considered high-order $g$~modes to probe the properties of convective cores in Slowly Pulsating B and $\gamma$~Dor stars.

%-------------------------------------------------------------------------
\section{\emph{Kepler}'s first observations}

\begin{figure}
\epsscale{1.1}
\plotone{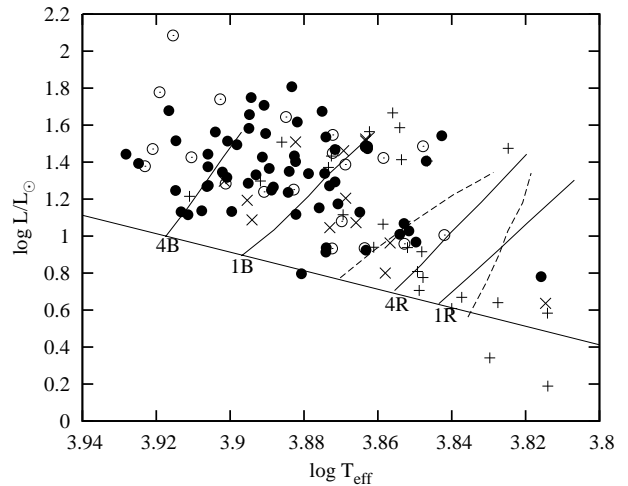}
\caption{Theoretical HRD showing classification of {\kepler} target stars. Filled circles are $\delta$\,Sct; open circles $\delta$\,Sct/$\gamma$\,Dor; crosses $\gamma$\,Dor/$\delta$\,Sct and plus signs $\gamma$\,Dor. The solid lines show the Zero-Age Main Sequence and the radial fundamental red and blue edges (1R, 1B) and the 4-th overtone radial red and blue edges (4R, 4B) \citep{Dupt05}. The dashed lines are the red and blue edges of the $\gamma$\,Dor instability strip ($\ell$ = 1 and mixing length $\alpha_{MLT}$ = 2.0) \citep{Dupt05}.}

 \label{fig:HRKepler}
\end{figure}

The main result from the inspection of this first {\kepler} data is that there are practically no pure $\delta$\,Sct or $\gamma$\,Dor pulsators. 
This finding raises a new problem: why is it that there is a clear distinction between the two types of pulsator in ground-based observations but not in {\kepler} observations?
Part of the answer could be that in ground-based photometric observations the modes of highest visibility are modes of low degree whereas at the photometric precision of {\kepler} many more modes of high degree are now visible.

Now we must consider whether we can construct a classification scheme that allows us to retain the useful $\delta$\,Sct/$\gamma$\,Dor categories. It is important to notice that relying only on the mode frequencies, {\kepler}'s data show that all stars with variability in these
frequency ranges are hybrids.  The use of amplitude in addition to frequency can help by
potentially filtering out the high-degree modes that generally would be expected to have
a lower photometric amplitude. 
Considering both amplitude and frequency, we propose a new observational classification scheme for $\delta$\,Sct and $\gamma$\,Dor stars:

\begin{itemize}

\item $\delta$\,Sct: most of the frequencies are $\ge$ 5 d$^{-1}$, and the lower frequencies are of relatively low amplitude;

\item $\delta$\,Sct/$\gamma$\,Dor hybrid: most of the frequencies are $\ge$ 5 d$^{-1}$, but there are some lower frequencies which are of comparable amplitude;

\item $\gamma$\,Dor: most of the frequencies are $\le$ 5 d$^{-1}$, and the higher frequencies are of relatively low amplitude.

\item $\gamma$\,Dor/$\delta$\,Sct hybrid: most of the frequencies are $\le$ 5 d$^{-1}$, but there are some higher frequencies which are of comparable amplitude; 

\end{itemize}

We apply this scheme to a sample of 554 stars selected from the {\kepler} working group targets, where we have excluded stars that are constant or eclipsing
variables. Among them, 234 show $\gamma$\,Dor or $\delta$\,Sct frequencies and have an
effective temperature near the instability strips for these types of
pulsation. Table 1
summarizes the categorization and mean properties for these stars.
The same stars are shown in the theoretical HRD in Fig.~\ref{fig:HRKepler} using the $\Teff$ and radii
derived from the {\kepler} Input Catalogue (KIC). Also shown are calculated red and blue edges of the instability strips \citep{Dupt05}. We see that there is some separation between categories, but it is not as clean as one might have hoped. We also see that the theoretical red and blue edges are not a good match for the observations.

\begin{table}
\begin{center}
\caption{Number of stars and percentage in each class. For each class the mean values of the effective temperature is given.}\label{tbl-1}
\begin{tabular}{lccc}
Class & Number & Percent & $<$ log $\Teff$ $>$ \\
\hline
$\delta$\,Sct                 & 67  & 27 & 3.885 $\pm$ 0.003 \\
$\delta$\,Sct/$\gamma$\,Dor & 32  & 14 & 3.883 $\pm$ 0.006 \\
$\gamma$\,Dor/$\delta$\,Sct & 19  &  9 & 3.868 $\pm$ 0.006\\
$\gamma$\, Dor                & 116 & 50 & 3.853 $\pm$ 0.005\\
\hline
\end{tabular}
\end{center}
\end{table}

%*************
\begin{figure}
\epsscale{1.2}
\plotone{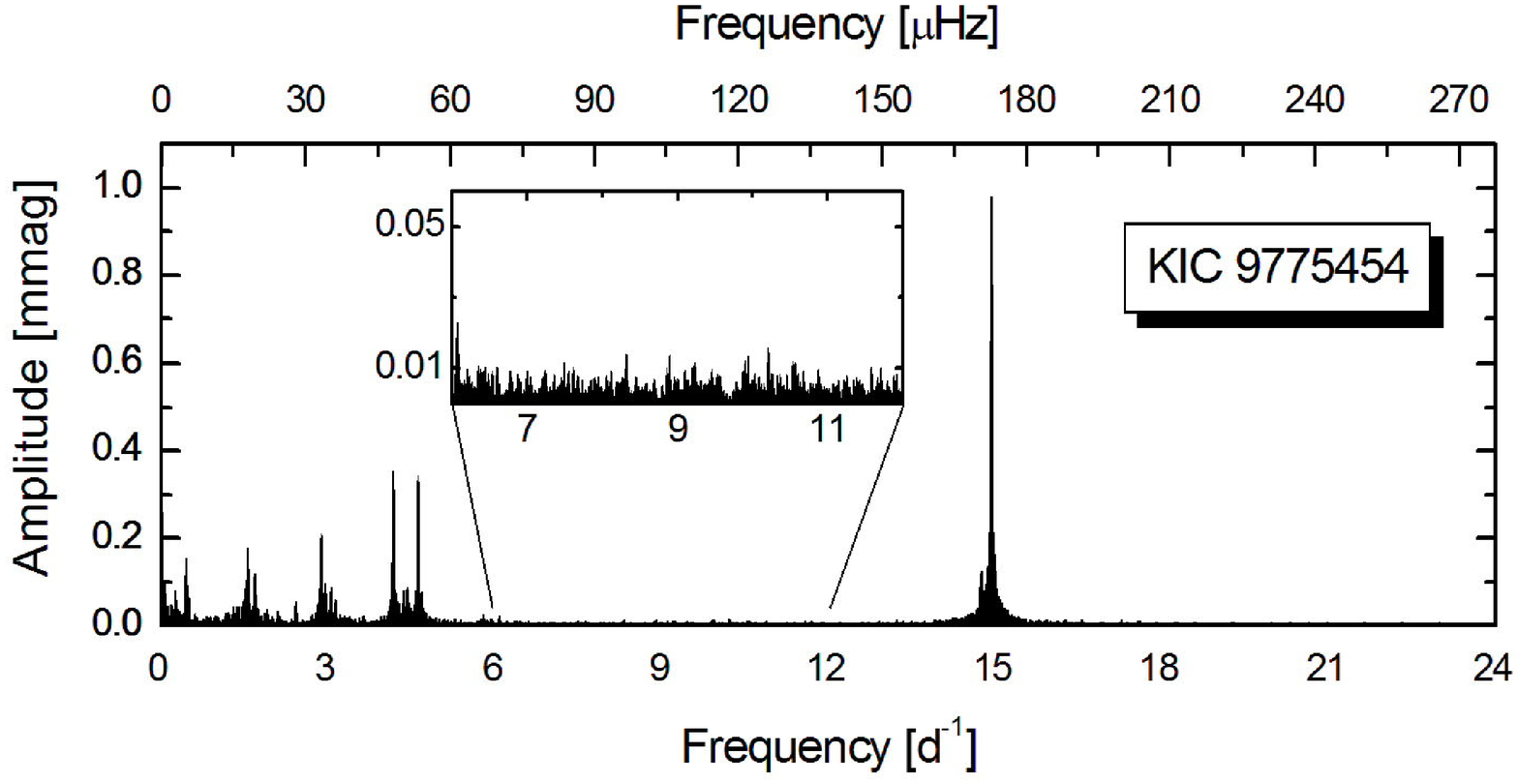}
\plotone{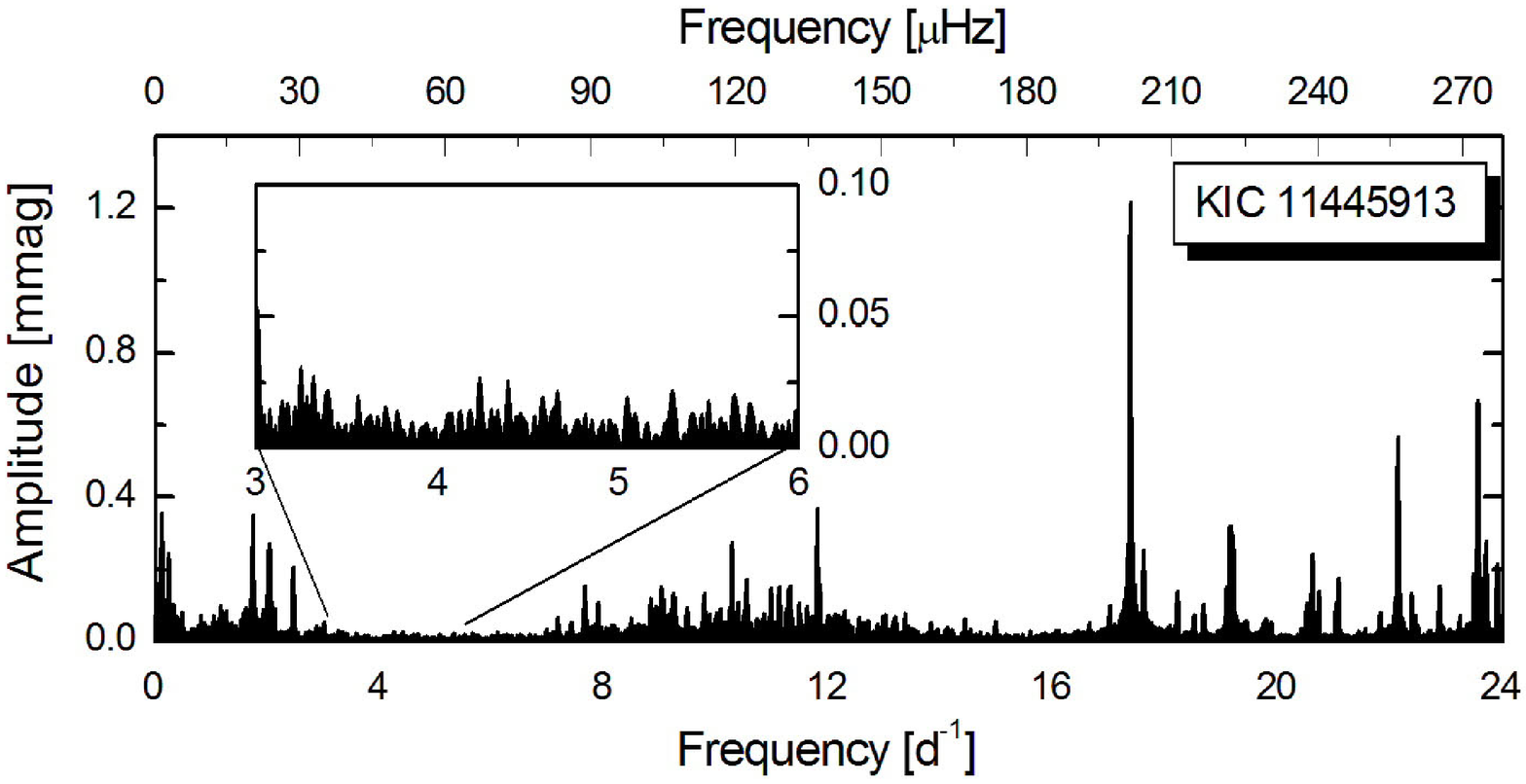}
\caption{Frequency spectra of the long-cadence data obtained by {\kepler} for the stars KIC~9775454 and KIC~11445913 that display hybrid pulsations in the $\gamma$\,Dor region (below 5\,d$^{{-}1}$) and in the $\delta$\,Sct region (above 12\,d$^{{-}1}$). 
The details of the spectra at higher frequencies require short-cadence data to be confirmed and analyzed.
\label{fig:kepler-obs}}
\end{figure}

The amplitude spectrum of two hybrid stars (\objectname{KIC~9775454} and \objectname{KIC~11445913}) are shown in Fig.~\ref{fig:kepler-obs}. 
The vertical line in Fig.~\ref{fig:instreg} indicates the position in effective temperature of the star \objectname{KIC~11445913}, confirming it as a potential hybrid candidate.
Fig.~\ref{fig:kepler-obs} also shows the very low level of noise measured in the gap between the frequency ranges for the $p$~modes and $g$~modes.
This gap, corresponding to a lack of unstable modes, is expected in the theory for hybrid pulsators.
However, in some stars observed by {\corot}, several frequencies can be found in the ``gap''.
The same behavior is also found for some of the {\kepler} targets.
An explanation for these frequencies still needs to be found from a more detailed study of the observations and by modeling the stars being observed. In some cases, rotational splitting seems to be the cause of modes in the frequency gap between $p$ and $g$ modes \citep{Boua09}.
Preliminary analysis of {\kepler} data for candidate hybrid stars gives very encouraging results for the number of hybrid stars with very high quality frequency spectra.
The large number of pulsational frequencies makes it possible to study global characteristics of the frequency spectrum such as asymptotic properties, and also individual frequency fittings.

%-------------------------------------------------------------------------
\section{Conclusions}

Our first look at the {\kepler} data provides encouraging results on the number of hybrid stars thanks to very high quality frequency spectra coupled with unprecedented photometric precision.
The {\kepler} data are essential to overcome the aliasing and photometric stability problems affecting the $\gamma$\,Dor frequency range associated with ground-based photometric campaigns.

The additional frequencies found in hybrid stars will help to overcome difficulties with mode identification and the detection of many $\gamma$\,Dor $g$~modes will enable analysis by asymptotic techniques.
The {\kepler} data on $\gamma$\,Dor and $\delta$\,Sct stars promise to help us to answer a number of outstanding questions:
why have {\kepler} (and {\corot}) detected a significant number of hybrids, whereas theory predicts the existence of hybrids in only a small overlapping region of the instability strips?
What is the connection between Am chemical peculiarity and hybrid behavior? 
Why are modes observed between the $\delta$\,Sct and $\gamma$\,Dor frequency ranges, where theory predicts a frequency gap?
Will we be able to confirm stochastic excitation or other pulsation driving mechanisms?
Can the high-photometric precision of {\kepler} detect higher-degree modes ($\ell{>}2$) that could fill in the frequency gap?
Will these stars help us to distinguish between explanations for the red edge of the $\delta$\,Sct instability strip?
Will we be able to discern rotational frequency splittings and derive internal differential rotation rates for these stars? 

Perhaps just as important are the many {\kepler} targets that lie within the $\gamma$\,Dor or $\delta$\,Sct instability strips that show no periodic variations at micromagnitude photometric precision levels obtainable by {\kepler}.
Monitoring over several years with {\kepler} will provide insight on amplitude limiting and mode selection mechanisms for these stars that is not possible with ground-based observations.

The analysis presented here of the first and second Quarter data obtained by {\kepler} confirms its capability to provide very high quality data that will allow us to study hybrid pulsators of the $\delta$\,Sct/$\gamma$\,Dor type. Moreover, the very rich frequency spectra of these stars leads us to propose a new classification scheme to overcome the ambiguity of the older, pure frequency-based scheme. The close frequency spacing of relatively short period modes in $\delta$\,Sct stars requires short-cadence data. We expect to obtain short-cadence data for selected targets of mixed character in the near future.

\vglue 2pt
%-------------------------------------------------------------------------
\acknowledgements
Funding for this Discovery mission is provided by NASA\textquoteright s
Science Mission Directorate. The authors gratefully acknowledge the entire {\kepler} team, whose outstanding efforts have
made these results possible. The authors also thank all funding councils and agencies that have supported the activities of KASC Working Groups~4 and 10.

%-------------------------------------------------------------------------
{\it Facilities:} \facility{The {\kepler} Mission}.

%-------------------------------------------------------------------------
\enddocument